\documentclass[aps,pra,reprint, amsmath, amssymb,superscriptaddress,nofootinbib]{revtex4-1}

\usepackage{bm}
\usepackage[retainorgcmds]{IEEEtrantools}
\usepackage{graphicx}
\usepackage{mathrsfs}
\usepackage{amsmath}
\usepackage{amssymb}
\usepackage{color}
\usepackage{amsfonts}
\usepackage{times,txfonts}
\usepackage{nicefrac}
\usepackage[colorlinks=true,linkcolor=blue,urlcolor=blue,citecolor=blue,pdfusetitle]{hyperref}
\usepackage{amsmath}

\DeclareMathOperator{\sech}{sech}

\begin{document}

\title{Thermodynamic variational relation}
\date{\today}
\author{Domingos S. P. Salazar}
\affiliation{Unidade de Educa\c c\~ao a Dist\^ancia e Tecnologia,
Universidade Federal Rural de Pernambuco,
52171-900 Recife, Pernambuco, Brazil}

\begin{abstract}
In systems far from equilibrium, the statistics of observables are connected to entropy production, leading to the Thermodynamic Uncertainty Relation (TUR). However, the derivation of TURs often involves constraining the parity of observables, such as considering asymmetric currents, making it unsuitable for the general case. We propose a Thermodynamic Variational Relation (TVR) between the statistics of general observables and entropy production, based on the variational representation of $f$-divergences. From this result, we derive a universal TUR and other relations for higher-order statistics of observables.
\end{abstract}
\maketitle{}


{\bf \emph{Introduction -}}
The second law of thermodynamics posits that entropy production is nonnegative. While the random nature of the entropy production, denoted as $\Sigma$, may seem insignificant in large-scale systems, it takes center stage at smaller scales due to the pronounced impact of thermal and quantum fluctuations. \cite{RevModPhys.93.035008,Campisi2011,Seifert2012Review,Campisi2011,Esposito2009,Ciliberto2013,Crooks1998,Crooks1999,Gallavotti1995,Evans1993,Hanggi2015,Batalhao2014,Pal2019,TanVanVu2023}.

In the realm of nonequilibrium thermodynamics, physical observables, such as particle currents, heat, and work, undergo fluctuations. These fluctuations are encapsulated within the probability density function (pdf) $p(\Sigma, \phi)$, where $\Sigma$ represents the entropy production and $\phi$ the underlying observable, where the pdf typically depends on both the system and time, particularly in transient regimes.

At the heart of these fluctuations is the detailed fluctuation theorem (DFT) \cite{Crooks1998,Evans2002,Gallavotti1995,Jarzynskia2008,Jarzynski1997}, which equates entropy production (or some generalization thereof) to the ratio of probabilities,
\begin{equation}
\label{DFT}
\Sigma(\Gamma):=\ln \frac{P(\Gamma)}{P(\Gamma^\dagger)},
\end{equation}
where $\Gamma$ is some process and $m(\Gamma)=\Gamma^\dagger$ is a conjugate (involution), such that $m(m(\Gamma))=\Gamma$. As a consequence of (\ref{DFT}), for instance, we have the integral FT, $\langle \exp(-\Gamma)\rangle =1$, and the second law, $\langle \Sigma \rangle \geq 0$, from Jensen's inequality. 
The broad applicability of the fluctuation theorem (\ref{DFT}), especially when systems are far from equilibrium, renders it a useful tool in the study of nonequilibrium physics. This utility has prompted extensive theoretical and experimental investigation of these theorems within classical systems. The relevance of the detailed fluctuation theorem extends to the quantum realm \cite{Esposito2009,Jarzynski2004a,Campisi2011,Hasegawa2019a,Deffner2016}, specially to account for heat exchange between quantum-correlated bipartite thermal systems \cite{Micadei2019}. This extension serves to emphasize the pervasive importance of the fluctuation theorem.

A consequence of (\ref{DFT}) is the Thermodynamic Uncertainty Relation (TUR) \cite{Barato2015A,Gingrich2016,Polettini2017,Pietzonka2017,Hasegawa2019, Timpanaro2019b,Horowitz2022,Potts2019,Proesmans2019,Liu2020,Salazar2022d}, which reads in this formalism,
\begin{equation}
\label{TUR}
\frac{\langle \phi^2 \rangle - \langle \phi \rangle^2}{\langle \phi \rangle^2} \geq \sinh^{-2}(\frac{g(\langle \Sigma \rangle)}{2}),
\end{equation}
valid for any asymmetric current, $\phi(\Gamma^\dagger)=-\phi(\Gamma)$, and any involution $m$, where $P'(\Gamma):=P(\Gamma^\dagger)$ and $D(P|P')=\sum_{\Gamma} P(\Gamma)\ln[P(\Gamma)/P'(\Gamma)]$ is the Kullback-Leibler (KL) divergence, and $g(x)$ the inverse of $h(x)=x\tanh(x/2)$, for $x\geq 0$.

However, in the derivation of TURs, the current is defined as an observable $\phi(\Gamma)$, bearing a particular parity under involution. This specific constraint prompts the question of whether a relationship exists between the statistics of general observables and the entropy production. Such general relations were subject of a recent studies \cite{Liu2020,Dechant2020,Ziyin2023},

Expanding on the idea that the average entropy production is equivalent to the Kullback-Leibler (KL) divergence, we extend this consideration to encompass any $f-$divergence, demonstrating that it too can be represented in terms of the statistics of $\Sigma$. We then leverage a theorem from information theory \cite{Polyanskiy2022} to elucidate our main result for any convex function $f$: 
\begin{equation}
\label{main}
\langle \phi-e^{-\Sigma}f^{*}(\phi)\rangle \leq \langle f(e^{-\Sigma})\rangle,
\end{equation}
where $f^{*}$ represents the Legendre transform of $f$. The averages are taken over $p(\Sigma,\phi)$, and $\phi$ is any observable within the effective domain of $f^{*}$ (ensuring that $f^{*}(\phi)$ is finite). This relation, referred to as the Thermodynamic Variation Relation (TVR), is rooted in the variational representation of $f-$divergences. Notably, the TVR does not impose any constraints on the parity of $\phi$, thereby rendering it a universally applicable relation. The choice of $f$ determines the specific nature of the relationships, as is demonstrated in the various applications. While the TVR does not have strict limitations for specific classes of observables (like asymmetric currents in TURs), it is a tight expression for general observables, as explained further in the formalism section.

As one example of application of the TVR (\ref{main}), for a particular choice of $f$, we obtain a universal TUR \cite{Ziyin2023},
\begin{equation}
\label{varTUR}
\frac{\langle \phi^2 \rangle - \langle \phi \rangle^2}{\langle \phi(1 - e^{-\Sigma})\rangle^2} \geq \frac{1}{\langle e^{-2\Sigma}\rangle-1},
\end{equation}
for any $\phi$ such that $\langle \phi(1 - e^{-\Sigma})\rangle\neq 0$. Note that $\langle \phi (1-e^{-\Sigma})\rangle=\langle \phi \rangle_P - \langle \phi \rangle_{P'}$, which physically means the difference of the averages in the forward and backward processes. 

The paper is organized as follows. First, we present the formalism to prove our main result. Then, we discuss the result and apply it for different choices of $f$: total variation case, the $\chi^2$ case yielding the universal TUR (\ref{varTUR}) for general observables, and the $\alpha-$divergence case, resulting on thermodynamic relations consisting of high-order statistics of observables.

{\bf \emph{Formalism -}}
Let $\Gamma \in S$ and $m:S\rightarrow S$ is any involution $m(m(\Gamma))=\Gamma$, with
$\Gamma^\dagger:=m(\Sigma)$. Let $P:S \rightarrow [0,1]$ be a probability function and $P'(\Gamma):=P(\Gamma^\dagger)$. We consider $P$ such that $P(\Gamma)=0 \rightarrow P(\Gamma^\dagger)=0$ for any $\Gamma \in S$ (absolute continuity). Let $\phi:S\rightarrow \mathbb{R}$ be any finite observable ($\sup_\Gamma|\phi(\Gamma)| < \infty$). The pdf $p(\sigma,\phi)$ is defined as
\begin{equation}
p(\Sigma,\phi):=\sum_\Gamma P(\Gamma)\delta(\phi(\Gamma)-\phi)\delta(\Sigma(\Gamma)-\Sigma),
\end{equation}
Such that the average of a function $F(\Sigma,\phi)$ is given by
\begin{equation}
\int F(\phi,\Sigma)p(\Sigma,\phi)d\Sigma d\phi = \sum_\Gamma P(\Gamma) F(\phi(\Gamma),\Sigma(\Gamma)),
\end{equation}
thus we use $\langle \rangle = \langle \rangle_P=\langle \rangle_p$ for the averages over $p(\Sigma,\phi)$ or $P(\Gamma)$ interchangeably, unless stated otherwise. Let $f:(0,\infty)\rightarrow \mathbb{R}$ be a convex function. Define the convex conjugate $f^{*}$ as the Legendre transformation,
\begin{equation}
\label{cc}
f^{*}(y) := \sup_x xy - f(x).
\end{equation}
From (\ref{cc}), we have Fenchel's inequality for any $x \in dom(f)$ and $y \in dom(f^*)$, we have
\begin{equation}
\label{Fenchel}
f(x)+f^{*}(y) \geq xy.
\end{equation}
Now let $y:=\phi(\Gamma^\dagger)$, where  $\phi:S\rightarrow effdom(f^{*})$ is an observable that takes values in the effective domain of $f^{*}$ (ie, $f^*(y)$ is finite), and let $x:=P(\Gamma^\dagger)/P(\Gamma)$ in (\ref{Fenchel}), which results in
\begin{equation}
\label{Fenchel2}
f\big(\frac{P(\Gamma^\dagger)}{P(\Gamma)}\big)+f^{*}(\phi(\Gamma^\dagger
)) \geq \phi(\Gamma^\dagger) \frac{P(\Gamma^\dagger)}{P(\Gamma)}.
\end{equation}
Finally, replacing $P(\Gamma^\dagger)/P(\Gamma)=\exp(-\Sigma(\Gamma))$ from (\ref{DFT}), multiplying (\ref{Fenchel2}) by $P(\Gamma)$ and summing over all $\Gamma$ it yields
\begin{equation}
\label{Fenchel3}
\sum_\Gamma [f(e^{-\Sigma(\Gamma)})+e^{-\Sigma(\Gamma)}f^{*}(\phi(\Gamma))]P(\Gamma) \geq \sum_\Gamma \phi(\Gamma)P(\Gamma).
\end{equation}
Reordering the terms in (\ref{Fenchel3}), we get
\begin{equation}
\langle \phi - e^{-\Sigma} f^{*}(\phi) \rangle \leq \langle f(e^{-\Sigma})\rangle,
\end{equation}
which is our main result (\ref{main}) for any observable with $\phi(\Gamma) \in dom(f^{*})$ for all $\Gamma$.

{\bf \emph{Discussion -}}
We based the derivation above on a stronger result from information theory for $f-$divergences that we review and discuss below.

First, we define $f-$divergence $D_f(P|Q)$ as follows: Let $f:[0,\infty)\rightarrow \mathbb{R}$ be a convex function, $f(1)=0$ and $lim_{x\rightarrow 0^+} f(x)=f(0)$. The $f-$divergence $D(P|Q)$, for $P$ absolute continuous with respect to $Q$, is defined as 
\begin{equation}
\label{fdiv1}
D_f(P|Q):=\sum_\Gamma f\big(\frac{P(\Gamma)}{Q(\Gamma)}\big)Q(\Gamma).
\end{equation}
The result is a variational representation for $f-$divergences \cite{Polyanskiy2022}, namely
\begin{equation}
\label{infotheorem}
D_f(P|Q)=\sup_{\phi \in dom(f^{*})} \langle \phi \rangle_P - \langle f^{*}(\phi) \rangle_Q.
\end{equation}
When (\ref{fdiv1}) applied to $(P,P')$, the involution property makes $D_f(P|P')$ symmetric, 
\begin{equation}
\label{property1}
D_f(P|P')=D_f(P'|P)=\langle f(e^{-\Sigma}) \rangle.
\end{equation}
We also have from the involution property
\begin{equation}
\label{property2}
\langle f^{*}(\phi)\rangle_{P'} = \langle e^{-\Sigma}f^{*}(\phi)\rangle_P.
\end{equation}
Combining (\ref{property1}) and (\ref{property2}) in (\ref{infotheorem}), we obtain (\ref{main}) as a consequence of (\ref{infotheorem}) applied to $(P,P')$. Actually, (\ref{infotheorem}) is stronger, which makes (\ref{main}) a tight bound for general $\phi$. If, however, one constrains $\phi$ to a class of observables (for instance, asymmetric, $\phi(\Gamma^\dagger)=-\phi(\Gamma)$), then the bound might become loose as it does not considered any property of $\phi$ on the derivation.

In the subsequent applications, we will focus on a particular $f(x)$ in (\ref{main}). We then calculate $\langle f(\exp(-\Sigma)) \rangle$ and the Legendre transform $f^{*}$ respective to the chosen $f$. Ultimately, we determine the specific form of the TVR corresponding to this particular case. We select examples of $f$ that have previously been examined in information theory \cite{Polyanskiy2022}, and we translate them into the language of thermodynamics.

{\bf \emph{Application - total variation}} The result of this application can also be derived using other methods, but it is a staple example of how to use the TVR (\ref{main}). Consider the convex function $f(x)=|x-1|/2$. Note that $f$ is not differentiable at $x=1$, but it is not required for the result (\ref{main}). Moreover, $f$ satisfies the conditions for the $f-$divergence (\ref{fdiv1}) and $D_f(P|Q)$ is called the total variation distance in this case. The Legendre transform (\ref{cc}) is given by
\begin{equation}
\label{cctv}
f^{*}(y)=y,
\end{equation}
for $|y|\leq 1/2$ and $f^{*}=\infty$, for $|y|> 1/2$. Within the effective domain of $f^{*}$, we have from (\ref{main})
\begin{equation}
\label{TVRtv}
\langle \phi - e^{-\Sigma}\phi \rangle \leq \langle |1-e^{-\Sigma}|/2\rangle,
\end{equation}
for $|\phi(\Gamma)|\leq 1/2$ for all $\Gamma$. We multiply both sides by $2M\geq0$ and redefine $2M\phi \rightarrow \phi$ to obtain
\begin{equation}
\label{TVRtv2}
\langle \phi(1 - e^{-\Sigma}) \rangle \leq M\langle |1-e^{-\Sigma}|\rangle,
\end{equation}
for any $sup_\Gamma |\phi(\Gamma)|\leq M$, which can be written for any bounded $\phi$ if you choose $M=\sup_{\Gamma} (|\phi(\Gamma)|):=|\phi|_{max}$. Additionally, if $\langle \phi(1-e^{-\Sigma}) \rangle \neq 0$, we obtain
\begin{equation}
\label{TVRtv3}
\frac{2|\phi|_{max}}{\langle \phi(1-e^{-\Sigma})\rangle} \geq \frac{2}{\langle|1-e^{-\Sigma}|\rangle} = \frac{1}{\Delta(P,P')}\geq \frac{g(\langle \Sigma \rangle)}{\langle \Sigma \rangle},
\end{equation}
where introduced the total variation $\Delta(P,P')=\sum_{\Gamma}|P(\Gamma)-P'(\Gamma)|/2$ and we used a previous result \cite{Salazar2021b}, $\Delta(P,P')\leq \langle \Sigma \rangle/g(\langle \Sigma \rangle)$, where $g$ is defined after (\ref{TUR}). We note that (\ref{TVRtv3}) has a similar form to the TUR (\ref{TUR}), but it considers $|\phi|_{max}$ instead of the variance of $\phi$, and it is valid for any bounded observable with $\langle \phi(1-e^{-\Sigma}) \rangle \neq 0$. On a side note, the bound (\ref{TVRtv3}) is also tight for asymmetric observables, $\phi(\Gamma^\dagger)=-\phi(\Gamma^\dagger)$, and saturated by a minimal current $\phi(\Gamma)\in\{\phi_{max},-\phi_{max}\}$, just like the TUR (\ref{TUR}).

{\bf \emph{Application - $\chi^2$ and universal TUR}}
Now we consider the convex function $f(x)=(x-1)^2$. It satisfies the conditions for $f-$divergences and $D_f(P|Q)$ is the $\chi^2-$divergence. The Legendre transform $f^{*}$ (\ref{cc}) is given by
\begin{equation}
f^{*}(y)=y+y^2/4,
\end{equation}
for any $y$. We also have $\langle f(e^{-\Sigma}) \rangle = \langle e^{-2\Sigma}\rangle -1$, where we used $\langle e^{-\Sigma} \rangle =1$ explicitly. Then, the TVR (\ref{main}) reads
\begin{equation}
\langle \phi - e^{-\Sigma}(\phi + \frac{\phi^2}{4}) \rangle \leq \langle e^{-2\Sigma}\rangle -1.
\end{equation}
Now we redefine $\phi \rightarrow a(\phi-\langle \phi e^{-\Sigma} \rangle)$ and maximize with respect to $a$ and obtain $a=2\langle \phi(1-e^{-\Sigma} \rangle/(\langle \phi^2 e^{-\Sigma}\rangle -\langle \phi e^{-\Sigma}\rangle^2)$. Finally, redefining $e^{-\Sigma}\phi \rightarrow \phi$, we get
\begin{equation}
\frac{\langle \phi (1-e^{-\Sigma})\rangle^2}{\langle \phi^2 \rangle - \langle \phi \rangle^2} \leq \langle e^{-2\Sigma} \rangle -1. 
\end{equation}
For the case $\langle \phi (1-e^{-\Sigma}) \rangle \neq 0$, we get a universal TUR
\begin{equation}
\label{varTURdemo}
\frac{\langle \phi^2 \rangle - \langle \phi \rangle^2}{\langle \phi(1 - e^{-\Sigma})\rangle^2} \geq \frac{1}{\langle e^{-2\Sigma}\rangle-1}.
\end{equation}
For the specific case of asymmetric observables, $\phi(\Gamma^\dagger)=-\phi(\Gamma)$, we note that 
\begin{equation}
\frac{\langle \phi^2 \rangle - \langle \phi \rangle^2}{\langle \phi\rangle^2} \geq \sinh^{-2}(\frac{g(\langle \Sigma \rangle)}{2}) \geq \frac{4}{\langle e^{-2\Sigma}\rangle-1},
\end{equation}
which is consistent with the fact that the TUR (\ref{TUR}) is tight for the constrained case of asymmetric observables, and (\ref{varTUR}) is derived for the general case.

{\bf \emph{Application - $\alpha$ divergences}} Consider $f_{\alpha}(x)=(x^\alpha - \alpha x - (1-\alpha))/(\alpha(\alpha-1))$, for $\alpha \in (-\infty,0) \cup (0,1)$ and $x \in [0,\infty)$. In this case, $D_f(P|Q)$ is called $\alpha-$divergence. The Legendre transform $f^{*}$ (\ref{cc}) is given by
\begin{equation}
\label{ccalpha}
f^{*}(y)=\frac{h(y)^\alpha -1}{\alpha},
\end{equation}
for $y \in (-\infty,1/(1-\alpha))$, where $h(y):=((\alpha-1)y+1)^{1/(\alpha-1)}$. We also have 
\begin{equation}
\label{falpha}
\langle f_\alpha(e^{-\Sigma})\rangle = \frac{\langle e^{-\alpha \Sigma}\rangle-1}{\alpha(\alpha-1)},
\end{equation}
and the TVR (\ref{main}) reads
\begin{equation}
\label{TVRalpha}
\langle \phi - e^{-\Sigma}\frac{(h(\phi)^\alpha-1)}{\alpha} \rangle \leq \frac{\langle e^{-\alpha \Sigma}\rangle-1}{\alpha(\alpha-1)},
\end{equation}
for $\phi$ such that $\phi(\Gamma) \in (-\infty,1/(1-\alpha))$ for all $\Gamma$. Now redefine $\phi \rightarrow ( \phi^{\alpha-1}-1)/(\alpha-1)$, which makes $h(\phi)^\alpha \rightarrow \phi^\alpha$, resulting in
\begin{equation}
\label{TVRalpha2}
\frac{1}{\alpha(1-\alpha)}-\frac{\langle\phi^{\alpha-1}\rangle}{1-\alpha}-\frac{\langle\phi^\alpha e^{-\Sigma}\rangle}{\alpha}\leq \frac{\langle e^{-\alpha \Sigma}\rangle-1}{\alpha(\alpha-1)},
\end{equation}
for $\inf_\Gamma \phi(\Gamma) \geq 0$. Redefining $\phi \rightarrow a\phi$ with the same domain and optimizing for $a$, we get $a=\langle \phi^{\alpha-1}\rangle/\langle e^{-\Sigma}\phi^\alpha\rangle$, which leads to
\begin{equation}
\label{TVRalpha3}
\frac{1}{\alpha(\alpha-1)}\Big(\frac{\langle \phi^{\alpha-1}\rangle^\alpha}{\langle\phi^\alpha e^{-\Sigma}\rangle^{\alpha-1}}-1\Big)\leq \frac{\langle e^{-\alpha \Sigma}\rangle-1}{\alpha(\alpha-1)},
\end{equation}
which, for $\alpha=-n <0$ (such that $\alpha(\alpha-1)>0$) and redefining $\phi \rightarrow \phi^{-1}$, results in
\begin{equation}
\label{TVRalpha4}
\frac{\langle |\phi|^{n+1}\rangle^{n}}{\langle e^{-\Sigma}|\phi|^{n}\rangle^{n+1}}\geq \frac{1}{\langle e^{n \Sigma}\rangle},
\end{equation}
where we introduced $|\phi|$ so that the result (\ref{TVRalpha4}) applies to all observables for $n>0$. Note that this application is a sort of TUR for high-order statistics of observables and the entropy production.

{\bf \emph{Application - Hellinger's case}}
As a particular case of (\ref{TVRalpha3}), consider $\alpha=1/2$. In this case, $D_f(P|Q)$ is the squared Hellinger's distance. We have from (\ref{TVRalpha3})
\begin{equation}
\label{Hellingers1}
\langle e^{-\Sigma}|\phi|^{1/2}\rangle\langle|\phi|^{-1/2}\rangle\geq \langle e^{-\Sigma/2}\rangle^2.
\end{equation}
As (\ref{Hellingers1}) is valid for all $\phi$, we redefine $|\phi|^{1/2}\rightarrow \exp(s\phi)$ for any $s\in \mathbb{R}$, resulting in
\begin{equation}
\label{Hellingers2}
\langle e^{s\phi - \Sigma}\rangle \langle e^{-s\phi}\rangle \geq \langle e^{-\Sigma/2}\rangle^2.
\end{equation}
We also have from a previous result \cite{Salazar2023a} for the moment generating function, $G(-1/2):=\langle \exp(-\Sigma/2) \rangle \geq \sech(g(\langle \Sigma \rangle/2)$, resulting in
\begin{equation}
\label{Hellingers3}
\langle e^{s\phi - \Sigma}\rangle \langle e^{-s\phi}\rangle \geq \langle e^{-\Sigma/2}\rangle^2 \geq \sech^2(\frac{g(\langle \Sigma \rangle)}{2})\geq e^{-\langle \Sigma \rangle}.
\end{equation}
In summary, the first inequality in (\ref{Hellingers3}) is given by the TVR (\ref{main}) and the second is given by the bound for the mgf of entropy productions. Note that the last one is straigthforward from Jensen's inequality, $\langle e^{s\phi - \Sigma}\rangle \langle e^{-s\phi}\rangle \geq \exp(\langle s\phi - \Sigma -s\phi\rangle)=\exp(-\langle \Sigma \rangle)$. We note that (\ref{Hellingers3}) is a lower bound for the product of two mgfs: $G_F(-s)=\langle \exp (-s\phi)\rangle_P$ and $G_B(s)=\langle \exp(s\phi)\rangle_{P'}$, where $P'$ is the backwards process, valid for any observable $\phi$, obtained from the TVR (\ref{main}).

{\bf \emph{Conclusions - }}
We explored a result from information theory that states that $f$-divergences have a variational representation in terms of the supremum of the observable statistics. Then, we expressed the $f$-divergence in terms of the statistics of entropy production and obtained a general relation.

Depending on the choice of $f$, different relations can be derived. As applications, we obtained a relation for bounded observables in terms of the total variation (\ref{TVRtv3}), a universal TUR (\ref{varTUR}), a high-order statistics relation (\ref{TVRalpha4}), and a lower bound for the product of two moment-generating functions (mgfs) (\ref{Hellingers3}).

The relation (\ref{main}) utilizes the detailed fluctuation theorem in the form of (\ref{DFT}), even when $\Sigma$ is not the actual entropy production. For instance, in situations with quantum correlations \cite{Micadei2019}, all the results in the paper remain valid, as long as we replace $\Sigma$ with the appropriate term that contains the actual entropy production as well as other quantum information terms. For that reason, we expect this result to be useful in several situations.

\bibliography{lib5}
\end{document}